%% file: ms.tex
\documentclass[useAMS,usenatbib]{mn2e}

\usepackage{latexsym}
\usepackage{amssymb}
\usepackage{graphicx}

\usepackage{subfigure}

\input{macros}
\title[General Relativistic Force-Free Electrodynamics: A New Code and
  Applications to Black Hole Magnetospheres]{General Relativistic
  Force-Free Electrodynamics: A New Code and Applications to Black
  Hole Magnetospheres}

\author[Jonathan C. McKinney]{Jonathan
  C. McKinney\thanks{E-mail:jmckinney@cfa.harvard.edu}\\ Institute for
  Theory and Computation, Harvard-Smithsonian Center for Astrophysics,
  60 Garden Street, MS 51, Cambridge, MA 02138, USA}

\begin{document}
\date{Accepted 2006 January 18. Received 2005 Dec 8; in original form
  2005 Dec 8}
\pagerange{\pageref{firstpage}--\pageref{lastpage}} \pubyear{2005}
\maketitle
\label{firstpage}

\begin{abstract}

The force-free limit of magnetohydrodynamics (MHD) is often a
reasonable approximation to model black hole and neutron star
magnetospheres.  We describe a general relativistic force-free
(GRFFE) formulation that allows general relativistic
magnetohydrodynamic (GRMHD) codes to directly evolve the GRFFE
equations of motion. Established, accurate, and well-tested
conservative GRMHD codes can simply add a new inversion piece of
code to their existing code, while continuing to use all the
already-developed facilities present in their GRMHD code.  We show
how to enforce the $\mathbf{E}\cdot\mathbf{B}=0$ constraint and
energy conservation, and we introduce a simplified general model of
the dissipation of the electric field to enforce the $B^2-E^2>0$
constraint.  We also introduce a simplified yet general method to
resolve current sheets, without much reconnection, over many
dynamical times.  This formulation is incorporated into an existing
GRMHD code (HARM), which is demonstrated to give accurate and robust
GRFFE results for Minkowski and black hole space-times.

\end{abstract}




\section{Introduction}\label{introduction}

General relativistic force-free electrodynamics (GRFFE) is the
low-inertia limit of general relativistic magnetohydrodynamics (GRMHD)
(see, e.g., \citealt{kom02b,kom04a}).  Neutron star and black hole
magnetospheres exhibit regions of space that are very nearly
force-free, and self-consistent moderately realistic quasi-analytic
solutions exist that describe the ideal MHD or force-free environment
of such systems (see, e.g.,
\citealt{bz77,cont99,goodwin04,bn05,gru05}).  It is generally
difficult to solve the GRFFE equations to find even stationary
solutions, except with simplified assumptions, that apply to
astrophysical systems. For example, despite recent progress in studies
of neutron star magnetospheres, no self-consistent analytic solution
considers general relativistic effects.  Also, the solution of
\citet{bz77} is the only self-consistent analytic force-free solution
that has a realistic Poynting jet.  Also, analytic solutions typically
assume stationarity and axisymmetry and so rarely address the global
or local stability of the solutions against time-dependent or
nonaxisymmetric modes or stability against reconnection in current
sheets if present.

One simple technique to seek stationary solutions and to study
time-dependent stability is to directly numerically integrate the
GRFFE equations of motion.  Both the GRMHD and GRFFE equations of
motion can be written as a set of conservation laws that can be
directly integrated.  Conservative numerical GRMHD methods, such as of
HARM \citep{gmt03}, use so-called ``primitive'' quantities
($\mathbf{P}$) to define so-called ``conserved'' quantities
($\mathbf{U}$), fluxes ($\mathbf{F}$), and source ($\mathbf{S}$)
terms.  The temporal integration is determined by solving the set of
``conservation'' equations, which can be written as
\begin{equation}
\frac{\partial U^p(\mathbf{P})}{\partial t} = -\frac{\partial F^{pi}(\mathbf{P})}{\partial
  x^i} + S^p(\mathbf{P}) ,
\end{equation}
where $p$ labels the conservation equation and $i$ is the spatial
index. The set of source terms ($S^p$) accounts for the connection
coefficients or other sources of mass-energy-momentum.  Thus
conservation is explicitly true as long as the source terms vanish.
For an axisymmetric, stationary metric and as written in HARM,
energy and angular momentum are explicitly conserved to machine
error because the source terms vanish.  HARM has been successfully
used to study black hole accretion flows, winds, and Poynting jets
\citep{gmt03,gsm04,mg04,mckinney05a,mckinney05b,mckinney05c}.

While GRFFE is mathematically just the no-inertia limit of GRMHD,
numerical truncation errors limit the use of GRMHD codes in evolving
systems with regions where the magnetic energy density greatly exceeds
the rest-mass energy (see, e.g., \citealt{mg04,kom04b,kom05a,kom05b}).
Conversely, the GRFFE equations of motion do not describe the
rest-mass motion along field lines or the thermal energy of the
particles.  So long as the inertia of particles remains negligible,
the GRFFE equations of motion properly describe the magnetic field
geometry and motion.

A GRFFE code can be used to study neutron star and black hole
magnetospheres. For example, the origin of the collimation and
stability of astrophysical jets remains unexplained. A GRFFE code
can be used to examine the origin of the collimation and the
stability of any model for a Poynting-dominated jet.

Existing jet solutions may not be stable and self-collimation of
isolated jets may not work
\citep{spr96,okamoto1999,okamoto2000,okamoto2003}. Collimation by a
predominately toroidal field (hoop stress) may lead to a pinch/kink
instability, whereas poloidal collimation works only if the jet is
surrounded by an extended disk wind
\citep{spr96,begli94,begelman1998}.  The solution to the collimation
problem may be that the disk wind collimates the Poynting jet.  By
using both GRMHD and GRFFE numerical models, the origin of
collimation, acceleration, and stability of jets can be examined.
Thus it is more efficient to have a single code be able to perform
both GRMHD and GRFFE studies.

Even a black hole system accreting a thick disk has a force-free
magnetosphere \citep{mg04}.  Such a magnetosphere quantitatively
agrees with the solution by \citet{bz77} for low black hole spins
and agrees qualitatively for high spins \citep{mg04,kom04b,kom05a}.
Gamma-ray bursts (GRBs), active galactic nuclei (AGN), and x-ray
binaries probably exhibit both a Poynting-dominated jet and a jet
from the disk, as suggested by GRMHD numerical models
\citep{mckinney05b,mckinney05c}. In particular, of all jet
mechanisms, the Blandford-Znajek driven jet is the only one that can
clearly produce an ultrarelativistic jet \citep{mckinney05a}. Thus
is it important to study the Blandford-Znajek process in detail.  A
GRFFE code can help determine the stability of the
Poynting-dominated jet generated by the Blandford-Znajek effect. For
example, a GRFFE code was used to study the pure monopole version of
the Blandford-Znajek solution, which is found to be stable
\citep{kom01,kom02a}.  However, general Poynting-dominated jets
produced by the Blandford-Znajek effect may be violently unstable to
pinch and kink instabilities (see, e.g., \citealt{li2000}, but see
also \citealt{tom01}).  A GRFFE code proves valuable in stability
analyses by avoiding the difficulty of analytically crossing the
light cylinder (see, e.g., \citealt{tom01}), which manifests itself
as a singularity in the Grad-Shafranov equation for stationary
solutions.

The goal of this paper is to formulate the GRFFE equations of motion
such that a conservative GRMHD code can be used with small
modifications. Present numerical methods directly evolve the
electric and magnetic fields (primitive quantities are
$\{\mathbf{E},\mathbf{B}\}$, the field-evolution approach) and have
to explicitly check that the velocity is less than the speed of
light (i.e. magnetic energy density is greater than electric energy
density, $B^2-E^2>0$) since the electric and magnetic field are
evolved with no constraint on their evolution \citep{kom02b,kras05}.
The method we describe evolves the drift velocity and magnetic field
(primitive quantities are $\{\mathbf{v},\mathbf{B}\}$,
(velocity-evolution approach), explicitly breaks down when that
constraint is violated and this ensures a physical and causal
evolution.

We show that this velocity-evolution approach can be directly used
by GRMHD codes, and explicitly discuss how this is done for
conservative numerical methods. Many GRMHD codes have recently been
developed \citep{koide99,kom99,gmt03,dh03,ss05,anninos05,anton05}.
Our discussion applies to any GRMHD method, but focuses on
conservative-based codes.  We show that a separate GRFFE code does
not have to be developed, and many of the tools and methods used to
make the GRMHD code perform well carry directly over to the GRFFE
version.

\S~\ref{forcefree} shows how a conservative GRMHD code can be used to
evolve the GRFFE equations of motion.  Simplified models of
dissipation in GRFFE are discussed in order to handle regions here the
electric field dominates the magnetic field.

\S~\ref{tests} includes a series of standard tests of the GRFFE code.

\S~\ref{models} discusses models of black hole magnetospheres.  This
section demonstrates the usefulness of the GRFFE equations of motion
and the model of dissipation.

\S~\ref{conclusions} summarizes the results of the paper.

The GRMHD notation follows \citet{mtw73} and we use Heaviside-Lorentz
units unless otherwise specified, which is like Gaussian units without
the $4\pi$.  For example, the 4-velocity components are $u^\mu$
(contravariant) or $u_\mu$ (covariant).  For a black hole with angular
momentum $J=j GM^2/c$, $j=a/M$ is the dimensionless Kerr parameter
with $-1\le j\le 1$.  The contravariant metric components are
$g^{\mu\nu}$ and covariant components are $g_{\mu\nu}$.  The comoving
energy density is $b^2/2$, where $b^\mu$ is the comoving magnetic
field.  See \citet{gmt03,mg04} for details.

\section{Force-Free Electrodynamics}\label{forcefree}

This section shows how a conservative-based algorithm designed for
solving the ideal GRMHD equations of motion based upon the velocity
and a magnetic field can be used to solve the GRFFE equations of
motion.  In particular, the electric and magnetic field are used to
derive the velocity of the frame in which the electric field vanishes.
This allows a conservative ideal GRMHD numerical code to evolve the
force-free equations of motion by simply modifying the inversion from
conserved quantities to primitive quantities.

The force-free electrodynamics equations of motion are the 8 Maxwell
equations:
\begin{equation}\label{firstmax}
\nabla_\mu F^{\mu\nu} = -J^\nu ,
\end{equation}
where $F^{\mu\nu}$ are the components of the Faraday tensor and
$J^\mu$ are the components of the current density, and
\begin{equation}
\nabla_\mu \dF^{\mu\nu} = 0 ,
\end{equation}
where $\dF^{\mu\nu}\equiv \frac{1}{2}
\epsilon^{\mu\nu\alpha\beta}F_{\alpha\beta}$ are the components of the
dual Faraday (or Maxwell) tensor, $F^{\mu\nu}\equiv -\frac{1}{2}
\epsilon^{\mu\nu\alpha\beta}\dF_{\alpha\beta}$,
$\epsilon^{\alpha\beta\delta\gamma} \equiv (-1/\detg)
[\alpha\beta\gamma\delta]$, and $\epsilon_{\alpha\beta\delta\gamma}
\equiv \detg [\alpha\beta\gamma\delta]$, where brackets denote the
unit length completely antisymmetric tensor.  These 8 equations define
6 evolution equations and 2 differential constraints.

By the antisymmetry and duality of the Faraday and Maxwell tensors,
they be expressed in component form as
\begin{equation}\label{faraday}
F^{\alpha\beta} \equiv f\eta^\alpha E^\beta - f\eta^\beta E^\alpha -
h B_\gamma \eta_\delta \epsilon^{\alpha\beta\gamma\delta}
\end{equation}
and
\begin{equation}
\dF^{\alpha\beta} \equiv -h\eta^\alpha B^\beta + h\eta^\beta B^\alpha -
f E_\gamma \eta_\delta \epsilon^{\alpha\beta\gamma\delta} ,
\end{equation}
where $f$ and $h$ are arbitrary independent constants that we set to
be $f=h=1$ consistent with the conventions in \citet{mtw73}, and see
also, e.g., \citet{bs03}.  Here $E^\alpha \equiv \eta_\beta
F^{\alpha\beta}$, $B^\alpha \equiv \eta_\beta \dF^{\beta\alpha}$, and
$\eta^\mu$ is an arbitrary 4-vector.  This gives that $F_{\mu\nu}
\dF^{\mu\nu} = 4E^\mu B_\mu (-\eta^2)$ and $F^{\mu\nu} F_{\mu\nu} =
2(B^2-E^2)(-\eta^2)$.  For a time-like $\eta^\mu$, the sign
conventions are as in special relativity.

The electromagnetic stress energy tensor is constructed from the
Faraday tensor and must be quadratic in the field strengths,
symmetric, and is divergenceless in vacuum by Maxwell's equations.  This
unique tensor is
\begin{equation}\label{tmunu}
T^{\mu\nu} = {F^\mu}_\lambda F^{\nu\lambda} - \frac{1}{4}g^{\mu\nu}
F^{\lambda\kappa} F_{\lambda\kappa} ,
\end{equation}
where $g^{\mu\nu}$ are the components of the metric.  The
duality between the Faraday and Maxwell tensors and the definitions of
$E^\mu$ and $B^\mu$ give that
\begin{eqnarray}\label{bigt}
T^{\mu\nu} & = & \left(\frac{B^2+E^2}{2}\right)\left(\eta^\mu \eta^\nu +
P^{\mu\nu}\right) \\
& - & \left(-\eta^2\right)\left(B^\mu B^\nu + E^\mu E^\nu\right)\nonumber \\
& - & \eta_\alpha E_\beta B_\kappa \left(\eta^\mu \epsilon^{\nu\alpha\beta\kappa}+\eta^\nu \epsilon^{\mu\alpha\beta\kappa}\right)\nonumber,
\end{eqnarray}
where the projection operator $P^{\mu\nu} = (-\eta^2)g^{\mu\nu} +
\eta^\mu\eta^\nu$.  For example, the energy density in frame
$\eta^\mu$ is $\eta_\mu \eta_\nu T^{\mu\nu} = \eta^4(E^2+B^2)/2$,
which for a time-like $\eta^\mu$ is the same expression as in
special relativity.  If the electromagnetic field is the only source
of stress-energy, then equations (\ref{firstmax}) are equivalent to
the energy-momentum conservation equations
\begin{equation}\label{emcons}
\nabla_\mu T^{\mu\nu} = J_\mu F^{\mu\nu} = 0 ,
\end{equation}
where only 2 of the 4 components of equation (\ref{emcons}) are
independent because equations (\ref{emcons}) implies
\begin{equation}
F_{\mu\nu} \dF^{\mu\nu} = 4E^\mu B_\mu (-\eta^2) = 0 .
\end{equation}
For more details see \citet{kom02b,kom04a}.

\subsection{Inversion}\label{inversion}

Conservative numerical GRMHD methods, such as of HARM, operate
primarily on so-called ``primitive'' quantities ($\mathbf{P}$): fluid
density, fluid internal energy, coordinate fluid velocity, and the
lab-frame coordinate magnetic field.  HARM uses the primitive
quantities to define so-called ``conserved'' quantities ($\mathbf{U}$)
and the fluxes ($\mathbf{F}$), which are both closed-form operations.
These conserved quantities can be evolved forward in time, but then an
inversion to primitive quantities is required to easily define the
fluxes and other required quantities to determine the next update.

No closed form solution appears to exist in GRMHD, so most inversion
methods rely on iterative procedures such as Newton's method. This
approach can be used in force-free electrodynamics, but may not
always work due to known problems (such as poorly conditioned or
singular Jacobians) with Newton's method.  However, a closed-form
solution does exist for force-free electrodynamics, as described
below.

The only conserved quantities of relevance in force-free
electrodynamics are the lab-frame momentums $T^t_\mu$ (the energy
evolution equation giving the $\mu=t$ term is actually redundant)
and the lab-frame magnetic field $B^\mu$ (only 3 components are
independent).  If one could obtain $E^\mu$, then one could
reconstruct the Faraday from equation (\ref{faraday}) or any other
quantities from $E^\mu$ and $B^\mu$.  It is straight-forward to show
that if $E^\alpha B_\alpha=0$, then
\begin{equation}\label{einv}
E^\alpha = \epsilon^{\alpha\beta\gamma\delta} B_\beta T^\xi_\gamma
\eta_\xi \eta_\delta / (B^2 \eta^4) ,
\end{equation}
as shown by a substitution of equation (\ref{tmunu}) and equation
(\ref{faraday}) into this expression.  Only the spatial components
of $\eta_\mu T^\mu_\nu$ and $B^\mu$ are needed if one chooses a
special form of $\eta^\mu$.  Notice that $E^\alpha B_\alpha = 0$ is
explicitly true, therefore the degeneracy condition of
$\mathbf{E}\cdot\mathbf{B}=0$ can be preserved to machine error
regardless of the truncation error in $\mathbf{T}$ or $\mathbf{B}$.
Also notice that equation (\ref{einv}) projects out a component
perpendicular to time, the spatial field, and the momentums.  For a
fixed field $B^i$, only 2 components of the electric field are
independent.

One may choose to have $\eta^2 = -1$ such that $\eta^\mu$ only
depends on the metric. One choice is $\eta_\mu=\{-\alpha,0,0,0\}$,
where $\alpha \equiv 1/\sqrt{-g^{tt}}$.  Then
$\eta^\mu=(1/\alpha)\{1,-\beta^i\}$, where $\beta^i\equiv \alpha^2
g^{ti}$. This defines a zero angular momentum (ZAMO) frame.  This
choice of $\eta^\mu$ makes it possible to only require $T^t_i$ and
$B^i$ to obtain $E^\alpha$ in equation (\ref{einv}).

Another interesting choice for $\eta^\mu$ is to have $E^\nu \equiv
\eta_\mu F^{\mu\nu} = 0$.  In the ideal GRMHD equations of motion,
for a fluid velocity $u^\mu$, this choice corresponds to the
electric field in the comoving frame $u^\mu$ being $e^\nu \equiv
u_\mu F^{\mu\nu} = 0$.  Then $\eta^2 = -1$ since the fluid velocity
is time-like.  In force-free electrodynamics, there is no unique
4-velocity that satisfies $e^\nu = 0$, but one such frame is
constructed below that is uniquely always time-like with a minimum
Lorentz factor with respect to the frame with 4-velocity $\eta^\mu$.

As shown next, any two frames with 4-velocities $u^\mu$ and
$\eta^\mu$ can be easily related to determine a 4-velocity of the
frame in which $e^\nu = 0$ for the Faraday defined in terms of
$\eta^\mu$. Thus the metric and Faraday alone determine a 4-velocity
that allows one to use the ideal GRMHD Faraday,
\begin{equation}\label{cofaraday}
F^{\alpha\beta} \equiv - b_\gamma u_\delta \epsilon^{\alpha\beta\gamma\delta}
\end{equation}
and Maxwell,
\begin{equation}\label{comaxwell}
\dF^{\alpha\beta} \equiv - u^\alpha b^\beta + u^\beta b^\alpha  ,
\end{equation}
where $b^\nu \equiv u_\mu \dF^{\mu\nu}$.  This formulation and sign
conventions are the same as used in HARM.  In HARM, a new 4-velocity
is introduced that is unique by being related to a physical observer
for any space-time and has well-behaved interpolated values,
\begin{equation}
\tilde{u}^i \equiv u^i -\gamma \eta^i ,
\end{equation}
where $\gamma=-u^\alpha \eta_\alpha$ and so $u^t = \gamma/\alpha$.
This additional term represents the spatial drift of the ZAMO frame
defined earlier.  One can show that $\gamma=(1+q^2)^{1/2}$ with
$q^2\equiv g_{ij} \tilde{u}^i\tilde{u}^j$.

To obtain the 4-velocity, notice that
\begin{equation}\label{ealpha}
0 = e^\alpha = -(v_\beta E^\beta)\eta^\alpha + E^\alpha + v_\beta B_\gamma \eta_\delta
\epsilon^{\alpha\beta\gamma\delta} ,
\end{equation}
where $v_\beta \equiv -u_\beta/(u^\alpha \eta_\alpha)$.  The most general
form of the 4-velocity that satisfies the above is
\begin{eqnarray}\label{vgeneral}
v_\beta = G\left(\frac{-\epsilon_{\beta\sigma\xi\eta}\eta^\sigma E^\xi B^\eta}{-\eta^2 B^2}\right) +
H \left(\frac{\eta_\beta}{-\eta^2}\right) \nonumber\\
+ K \left(\frac{E_\beta}{\sqrt{-\eta^2 E^2}}\right)+L \left(\frac{B_\beta}{\sqrt{-\eta^2 B^2}}\right)  ,
\end{eqnarray}
where $G$, $H$, $K$, and $L$ are functions to be determined.  Each
term represents one of four orthogonal directions, and when
multiplied by arbitrary functions represent the most general
solution. The only nontrivial term is the antisymmetric product
between the last term in equation (\ref{ealpha}) and the first term
in equation (\ref{vgeneral}). Substitution of this $v^\beta$ into
equation (\ref{ealpha}) gives that $G = 1$.  Since $v_\beta
\eta^\beta = -1$, then $H = 1$.  With $u_\beta = \gamma v_\beta$,
then $u^2 = -1$ gives that $\gamma = \sqrt{-\eta^2
B^2/((1-K^2)B^2-E^2)}$, which simply defines $\gamma = -u^\alpha
\eta_\alpha$ as the Lorentz factor between the two frames. All terms
proportional to $K$ are orthogonal to each other and to $E^\alpha$,
and so in general $K=0$. Hence, $u^\alpha E_\alpha = 0$. To
determine $L$, notice that $b^\alpha$ in $\dF^{\mu\nu}$ in equation
(\ref{comaxwell}) can be written as
\begin{equation}
b^\alpha = \frac{P^\alpha_\beta B^\beta}{\gamma} ,
\end{equation}
where $P^\alpha_\beta=\delta^\alpha_\beta+u^\alpha u_\beta$ is the
projection tensor.  Thus $b^\alpha$ has terms proportional to
$B^\alpha$ and $u^\alpha$.  So extra terms added to $u^\alpha$
proportional to $B^\alpha$ vanish due to the antisymmetry of the
Maxwell and so do not contribute to the stress-energy tensor or the
equations of motion.  Thus, the function $L$ parameterizes the
arbitrary velocity component along a field line, and we choose $L=0$
to minimize the Lorentz factor of the frame.  Hence, $u^\alpha
B_\alpha = 0$ and so $b^\alpha = B^\alpha/\gamma$.  With this choice
of $L$, if $B^2-E^2>0$, then the frame with $u^\mu$ is always
time-like.  Thus the frame defined by the 4-velocity
\begin{equation}\label{uinv}
u^\alpha = \left(\sqrt{\frac{B^2}{B^2-E^2}}\right) \left(\eta^\alpha -
\frac{\epsilon^{\alpha\beta\gamma\delta} \eta_\beta E_\gamma B_\delta}{B^2}\right)
\end{equation}
is time-like for any force-free electrodynamic solution.  Thus, by
construction, we have shown that in force-free electrodynamics that it
is possible to boost into a time-like frame where the electric field
vanishes and thus the Poynting flux vanishes (see also,
e.g. \citealt{kom02b}).  This also shows that force-free
electrodynamics is a causal limit of GRMHD as long as
$B^2-E^2>0$. This 4-velocity also represents a unique covariant
definition of the ``field-line velocity,'' and also describes the
field-line velocity even in ideal MHD.

A GRMHD code may only need the coordinate lab-frame 3-velocity.  Since
$u^t = \gamma \eta^t$, then for the earlier defined ZAMO frame
$\eta^\mu$, the coordinate lab-frame 3-velocity is given by
\begin{equation}\label{vinv}
v^i \equiv \frac{u^i}{u^t} = -\beta^i + \alpha^2 \frac{[ijk] E_j B_k}{\detg B^2}
\end{equation}
The second term in equation (\ref{uinv}) represents the purely
spatial ``$\mathbf{E}\times\mathbf{B}$ drift.''  Note that there is
no evolutionary constraint on $T^{\mu\nu}$ that forces $B^2-E^2>0$,
and when this is violated the force-free model is no longer
physical. The value of $v^\phi$ coincides with the ``field rotation
frequency'' $\Omega_F=F_{tr}/F_{r\phi}=F_{t \theta}/F_{\theta\phi}$
for stationary axisymmetric flows for which $v^r=v^\theta=0$.

Now the inversion from $\{T^t_i,B^i\}\rightarrow \{E^i,B^i\}$ and
then $\rightarrow \{v^i,B^i\}$ completely defines equations
(\ref{cofaraday}) and (\ref{comaxwell}) for a general relativistic
force-free electrodynamics evolution using a conservative GRMHD
code.

Obviously this formulation can be also used to study special
relativistic models as well.  Notice that in special relativity that
the derived velocity expression reduces to the so-called
``electromagnetic 3-velocity'' $\mathbf{v} =
\mathbf{E}\times\mathbf{B}/B^2 = \mathbf{S}/B^2$, where $\mathbf{S} =
\mathbf{E}\times\mathbf{B}$ is the Poynting flux.

As used in HARM, this formulation preserves $\nabla\cdot\mathbf{B}=0$
and $\mathbf{E}\cdot\mathbf{B}=0$ to round-off error for both the
GRMHD and GRFFE equations of motion.  Notice that this differs from
other formulations that only preserve $\nabla\cdot\mathbf{B}=0$ and
$\mathbf{E}\cdot\mathbf{B}=0$ to truncation error
\citep{kom02b,kom04a}

The fact that the field-evolution approach does not break down might
be considered an advantage when seeking stationary solutions.  In such
a case the evolution may have regions that only transiently have
$B^2-E^2<0$ and the integration can pass smoothly through this region
into a physical solution space.  A related advantage of the
field-evolution approach is that Runge-Kutta temporal evolution can
recover the correct temporal order of accuracy.  That is, without
characteristic interpolation, HARM uses Runge-Kutta to time step to
achieve higher order temporal accuracy.  Runge-Kutta is only first
order accurate for the first substep, but after all substeps are
completed, the method is accurate to arbitrary order.  The
velocity-evolution approach can yield an unphysical result for a
substep and be unable to continue or treat the result as a violation
of force-free electrodynamics, while the field-evolution approach can
avoid such first order errors and recover to arbitrary order accuracy.
However, we have found the velocity-evolution approach to be
sufficient.  This issue of falling outside the light cone is the same
issue one encounters when evolving the GRMHD equations of motion, and
in that case the author knows of no method that does not require the
velocity at some point during the integration, so the issue is treated
as a generic one that simply requires a more accurate integration.

\subsection{Currents}

This section shows that the currents can be computed without time
derivatives, which is numerically convenient to avoid storing data at
previous times.  In ideal MHD or force-free electrodynamics there are
many dependent ways of equally describing the same physics.
Researchers often invoke, such as in discussions of current closure,
the current and the magnetic field as a more intuitive set of
quantities than the electromagnetic fields.  The current could be
computed from
\begin{equation}\label{Jcurrent}
J^\alpha \equiv F^{\alpha\beta}_{;\beta} ,
\end{equation}
which apparently requires time derivatives.  In force-free
electrodynamics, however, the current (like the 4-velocity) sits in
the null space of $F^{\alpha\beta}$, i.e. $J^\alpha F^\beta_\alpha=0$.
Hence, $J^\alpha E_\alpha=0$.  The same analysis as in
section~\ref{inversion} leads to the same result for $J^\alpha$ except
the function $L$ is no longer arbitrary, where here
\begin{equation}
L = \left(\frac{J_\beta
  B^\beta}{\rho_{q,\eta}}\right)\left(\sqrt{\frac{-\eta^2}{B^2}}\right) ,
\end{equation}
where $\rho_{q,\eta} \equiv J^\alpha \eta_\alpha$, which only
actually requires spatial derivatives if
$\eta_\mu=\{-\alpha,0,0,0\}$. Plugging equations (\ref{faraday})
into equations (\ref{Jcurrent}) and contracting $J_\beta$ with
$B^\beta$ gives
\begin{equation}\label{jdotb}
J_\beta B^\beta = B^\beta\left(E^\alpha \eta_{\beta;\alpha}-\eta^\alpha
E_{\beta;\alpha} - (B_\gamma \eta_\delta
{{\epsilon_\beta}^{\alpha\gamma\delta}})_{;\alpha}\right) ,
\end{equation}
where the first and last terms only require spatial derivatives with
the chosen $\eta_\mu$.  Using
\begin{equation}
\dF^{\alpha\beta}_{;\beta}=0 ,
\end{equation}
and contracting with $E_\alpha$, the second term in equation
(\ref{jdotb}) can be written as
\begin{equation}
-B^\beta \eta^\alpha E_{\beta;\alpha} = B^\alpha E^\beta
\eta_{\beta;\alpha} - E_{\beta;\alpha} E_\gamma \eta_\delta
\epsilon^{\beta\alpha\gamma\delta} .
\end{equation}
Now, for the chosen $\eta_\mu$, each term in equation (\ref{jdotb})
only requires spatial derivatives.  So the current can be written,
only actually requiring spatial derivatives, as
\begin{equation}\label{Jfinal}
J^\alpha = \rho_{q,\eta} \left(\eta^\alpha -
\frac{\epsilon^{\alpha\beta\gamma\delta} \eta_\beta E_\gamma
B_\delta}{B^2}\right) + B^\alpha \left(\frac{J^\beta B_\beta}{B^2}\right) ,
\end{equation}
where finally
\begin{eqnarray}\label{jdotb2}
J^\beta B_\beta & = & E^\alpha B^\beta \left(\eta_{\beta;\alpha}+\eta_{\alpha;\beta}\right) \nonumber \\
& + &
\left(B_{\alpha;\beta}B_\gamma-E_{\alpha;\beta}E_\gamma\right)\left(\eta_\delta\epsilon^{\alpha\beta\gamma\delta}\right)
,
\end{eqnarray}
where the asymmetry between $E^\alpha$ and $B^\alpha$ just expresses
the asymmetry in Maxwell's equations for $J^\alpha$.  One can use
equation (\ref{uinv}) to write
\begin{equation}\label{Jfinal2}
J^\alpha = u^\alpha\left(\frac{\rho_{q,\eta}}{\gamma}\right)
+ B^\alpha \left(\frac{J^\beta B_\beta}{B^2} \right) .
\end{equation}
That is, there is a perpendicular drift current ($J_\perp$) and a
field-aligned current ($J_\parallel$), i.e.
\begin{equation}\label{Jexplain}
J^\alpha = J_\perp^\alpha + J_\parallel^\alpha .
\end{equation}
Equation (\ref{Jfinal2}) is Ohm's law in dissipationless GRFFE.  In
the special relativistic regime this reduces to a lab-frame current
density of
\begin{equation}
\mathbf{J} = \frac{\mathbf{E}\times\mathbf{B}(\nabla\cdot\mathbf{E}) +
  \mathbf{B}(\mathbf{B}\cdot(\nabla\times\mathbf{B})-\mathbf{E}\cdot(\nabla\times\mathbf{E}))}{B^2} ,
\end{equation}
where $\eta_\alpha J^\alpha = \rho_{q,\eta} = -\alpha J^t =
-\nabla\cdot\mathbf{E}$ is the charge density in the frame moving with
4-velocity $\eta^\mu$.  As in general, the special relativistic
equation obviously has no time derivatives.

\subsection{Jump Conditions}\label{jump}

Conservative schemes are often based upon 1D piece-wise constant
Godunov methods that achieve higher than first order accuracy by
relying on a one-dimensional interpolation from, e.g., grid cell
centers to cell interfaces.  The interpolated states are assumed to be
approximately constant over the cell face, otherwise a generalized
Riemann problem must be solved (see, e.g. \citealt{toro99}).  The 1D
Godunov scheme then explicitly treats the cell interface discontinuity
appropriately and reduces to a trivial form for smooth flows.
However, while the hydrodynamic equations of motion allow arbitrary
left and right initial states, the electromagnetic field must obey
general well-known jump conditions as a manifestation of the Bianchi
identities and the antisymmetry of the Faraday tensor (see, e.g.,
chpt. 15 of \citealt{mtw73}).  Alternatively stated, electrodynamics
can be written as a set of Hamilton-Jacobi equations involving the
vector potential and the electric field.  Arbitrary interpolation of
arbitrary quantities does not generally enforce these jump conditions.
Thus, these jump conditions must be self-consistently enforced by
interpolating appropriate quantities in the appropriate way.

HARM directly operates on the magnetic field rather than the magnetic
vector potential, so the scheme must enforce the electrodynamic jump
conditions on the electric and magnetic fields.  For schemes that only
interpolate in space and not time, one only requires the spatial jump
conditions.  The jump conditions across a one-dimensional space-like
surface for a single lab-frame time ($t$) are obtained by integrating
Maxwell's equations.  For simplicity define $\detg \tilde{E}_\alpha
\equiv t^\beta F_{\alpha\beta}$ and $\detg \tilde{B}_\alpha \equiv
t^\beta \dF_{\beta\alpha}$.  Also, consider three arbitrary orthogonal
space-like vectors $A,B,C$ that describe a space-like volume and the
time-like (often but not always Killing) vector
$t^\alpha=\{1,0,0,0\}$. For the homogeneous Maxwell equations,
\begin{equation}\label{intcurlE}
0 = \int\frac{(\detg \dF^{\mu\nu})_{,\nu}}{\detg} d\Sigma_\mu ,
\end{equation}
where $\dF^{ij} = [ijk]\tilde{E}_k$.  First, let $d\Sigma_\mu =
\epsilon_{\mu\alpha\beta\gamma} t^\alpha dA^\beta dB^\gamma$ be a
1-form 2-volume for a single lab-frame time (see, e.g.,
\citealt{lich67,lich76,anile}).  Then equation (\ref{intcurlE})
gives for the two spatial parallel components that
\begin{equation}\label{epara}
\Delta_C \int\detg \tilde{E}_A d_A  [CAB] = 0 ,
\end{equation}
where the $C$-direction is chosen as perpendicular to the surface and
never is a sum implied for $[CAB]$, which only gives the 3-signature
for arbitrary, but fixed, $A,B,C$ corresponding to any 3 spatial
directions.  Choosing instead $d\Sigma_\mu =
\epsilon_{\mu\alpha\beta\gamma} dA^\alpha dB^\beta dC^\gamma$, one has
for the contravariant field that
\begin{equation}\label{bperp}
\Delta_C \int \detg B^C dA dB  = 0 .
\end{equation}

Likewise, for the inhomogeneous Maxwell equations,
\begin{equation}\label{intcurlB}
\int J^\mu d\Sigma_\mu = \int\frac{(\detg F^{\mu\nu})_{,\nu}}{\detg}
d\Sigma_\mu ,
\end{equation}
where $F^{ij} = [ijk]\tilde{B}_k$.  With $d\Sigma_\mu =
\epsilon_{\mu\alpha\beta\gamma} t^\alpha dA^\beta dB^\gamma$, then
\begin{equation}\label{bpara}
\Delta_C \int \detg \tilde{B}_A dA = [CAB] \int \detg J^B dA dC ,
\end{equation}
for a possible surface current $J^B\equiv \delta(C) K^B$, where
upper (lower) $A,B,C$ denotes the contravariant (covariant)
components parallel to the surface. Choosing instead $d\Sigma_\mu =
\epsilon_{\mu\alpha\beta\gamma} dA^\alpha dB^\beta dC^\gamma$ for
space-like orthogonal $A$, $B$, and $C$, one has for the
contravariant field that
\begin{equation}\label{eperp}
\Delta_C \int \detg E^C dA dB = \int \eta_\mu J^\mu \detg dA dB dC ,
\end{equation}
for a possible surface charge $\rho_{q,\eta} = \eta_\mu J^\mu \equiv
\delta(C) \sigma_{q,\eta}$, where $\sigma_{q,\eta}$ is the surface
charge in the frame moving with 4-velocity $\eta^\mu$ and the
$C$-direction is taken to be across the surface.

For an infinitesimal surface $\detg dA dB$, lengths $\detg dA$ and
$\detg dB$, or for point values of the fields, these expressions
reduce to
\begin{equation}\label{epara2}
\Delta_C \tilde{E}_A  [CAB] = 0 ,
\end{equation}
\begin{equation}\label{bperp2}
\Delta_C B^C = 0 ,
\end{equation}
and
\begin{equation}\label{bpara2}
\Delta_C \tilde{B}_A = [CAB] K^B ,
\end{equation}
\begin{equation}\label{eperp2}
\Delta_C E^C = \sigma_{q,\eta} ,
\end{equation}
which apart from the concern with $\detg$, the contravariant
vs. covariant components, and $\tilde{E},\tilde{B}$ vs. $E,B$ is
identical to the Cartesian special relativistic expressions for the
jump conditions.  Notice that no distinction is made between bound and
free charges or currents, so these expressions are generally true.

The equations (\ref{epara2}) through (\ref{eperp2}) must be
preserved at the 1-D cell interfaces.  For centered schemes the
continuities can be enforced by using the same interpolation stencil
for the left/right interpolations to the cell interface from the
left/right cell centers. This is the method chosen for the GRFFE
version of HARM.  These constraints on the fields form an implicit
constraint on the drift velocity and the magnetic field.  Notice
that equations (\ref{bpara2}) and (\ref{eperp2}) enforce no specific
constraint unless there is an enforced surface charge or surface
current.  Also notice that these constraints must also be preserved
in ideal GRMHD.

Interpolating the electric and magnetic field can lead to unphysical
interface states if $B^2-E^2<0$ is unconstrained at the interface.  To
avoid this, one can interpolate $\gamma \tilde{E}_\alpha$ and $\gamma$
separately and then reconstruct $\tilde{E}_\alpha$ for a given
interpolated $\gamma$ at the interface.  This guarantees that the
interface state is physical and reasonably similar to the center
states.

An alternative scheme can be designed with a staggered grid that
automatically enforces these constraints \citep{luca2003}.  However,
both methods involve the same number of interpolations since in their
case they must interpolate the interface field components to the
center before performing the inversion from conserved to primitive
quantities.  They must also use a stencil that guarantees no
discontinuities at their cell center.  Thus both methods are
equivalent.

\subsection{Energy Conservation}

The formulation above for the inversion in force-free electrodynamics
only explicitly requires $T^t_i$ and not $T^t_t$, which is a similar
feature to other methods \citep{kom02b,kras05}.  Since the truncation
error in each $T^t_\mu$ is independent, the conserved quantity
associated with energy conservation ($T^t_t$) is generally
inconsistent with $T^t_i$.  Hence, energy is only conserved to
truncation error.  As for any numerical scheme that only conserves
energy to truncation error, this error can be used to gauge the
reliability of the integration.  However, in steady-state problems
this truncation error may be secular and build-up and lead to
unrealistic solutions.  It is fruitful to formulate the inversion to
enforce energy conservation and compare the nonconservative
integration with a conservative one to gauge the actual effect of the
truncation error.

One requires a solution for $T^t_i$ as a function of an arbitrary set
of three components of $T^t_\mu$.  Then an arbitrary choice can be
made to move the truncation error from energy to a momentum.  For
axisymmetric, stationary space-times the natural choice of momenta are
the radial and $\theta$ momenta.  This relationship between $T^t_\mu$
that only depends otherwise on $B^\mu$ can be obtained from
\begin{equation}\label{ET}
T^\mu_\nu \eta_\mu \eta^\nu = \eta^4\frac{E^2+B^2}{2} ,
\end{equation}
where equation (\ref{einv}) also gives that
\begin{equation}\label{esq}
E^2 = \left(\frac{-\eta_\alpha\eta_\beta T^\alpha_\gamma
  T^\beta_\delta}{B^2\eta^6}\right)\left(g^{\gamma\delta} -
  \frac{\eta^\gamma\eta^\delta}{\eta^2} - \frac{B^\gamma
  B^\delta}{B^2}\right) .
\end{equation}
Equation (\ref{esq}) does not actually depend on $T^t_t$ for our
choice of $\eta^\mu$.  Equation (\ref{ET}) gives $T^t_t$ in terms of
$T^t_i$ and $B^i$.  Notice that contracting with only $\eta^\nu$
shows that $\eta^\nu T^\mu_\nu$ just reduces to $\eta_\alpha B^\beta
T^\alpha_\beta=0$ for $\nu\neq t$.  This shows that of the three
$T^t_i$'s that only two are independent for a fixed $B^i$.  However,
for an arbitrary magnetic field, the evolution of all three
$T^t_i$'s is required to avoid singular expressions.

To replace any particular $T^t_i$ with $T^t_t$, one solves the
quadratic equation (\ref{ET}) for that spatial component in terms of
$T^t_t$ and the remaining spatial components.  This new set of
effective $T^t_i$'s can then be plugged into equation (\ref{einv}).
Notice that in general the quadratic equation gives two solutions.
This degeneracy is introduced by using the energy, which lacks
directional information.  In practice it is sufficient to use the
solution closest to the one obtained originally from integration of
only the spatial parts ($T^t_i$).  This procedure can be used to
keep energy and angular momentum conserved to machine error, unlike
in prior methods \citep{kom02b,kras05}.  Care must be taken for
numerical methods with a large truncation error in the energy, since
the evolution of the spatial and temporal stress-energy components
may be disparate.  The origin of this larger truncation error is the
larger nonlinearity of the energy compared to the momentums.  Lack
of energy-momentum conservation can also be due to dissipative
processes, as described in the next section.

\subsection{Dissipation in Force-Free Electrodynamics}\label{diss}

There is no evolutionary constraint in dissipationless force-free
electrodynamics that preserves $B^2-E^2>0$, whose violation is taken
as evidence that the plasma being modelled would have a nonnegligible
inertial back-reaction on the electric field.  This typically occurs
in current sheets or in regions where the inertia would restrict the
field to have a velocity of $v<c$.  A physical system responds by
dissipating the electric field into other forms of energy.

The dissipation of the electric field is determined by an Ohm's law.
For magnetospheres with an ample supply of charges (i.e. not
``charge-starved'') the Ohm's law in force-free electrodynamics is
well-approximated by a large conductivity along the magnetic field
and a vanishing conductivity perpendicular to the magnetic field
\citep{kom04a,kom05b}.  This reduces Ohm's law to the condition
$\mathbf{E}\cdot\mathbf{B}=0$ and the only perpendicular current is
the drift current with velocity given by equation (\ref{vinv}).

To model this dissipation, \citet{kom05b} use the prescription that if
$B^2-E^2<0$, then they introduce a large cross-field conductivity.
They also have to modify the drift velocity to keep $v<c$.  The
problem with this prescription is that there is no dissipation until
$v>c$.  Indeed, in currents sheets the rate of dissipation is related
to the drift velocity and is allowed to be $v\rightarrow c$, leading
to the fastest possible reconnection rate.

\citet{lyubarsky05} study the relativistic Sweet-Parker sheet and
Petschek configuration and determine that the rate of reconnection is
much less than the speed of light, contrary to previous estimates
(see, e.g. \citealt{lyutikov03,lu03}).  This suggests that dissipation
should limit the drift of field into a current sheet.

First, inertial losses due to the dissipation of the electric field
in a relativistic flow, such as beyond the light cylinder of a
rotating compact object, can be ``captured'' by limiting the Lorentz
factor of the drift velocity.  This is a similar approach taken in
GRMHD numerical models in order to avoid significant numerical
errors at large Lorentz factors.  A simple prescription is to limit
$\gamma\equiv -u^\alpha \eta_\alpha$ such that
$1\le\gamma^2\le\gamma_{max}^2$.  From $-1=u^\alpha u_\alpha$, this
means replacing $B^2$ in equation (\ref{vinv}) (and the similar
$B^2$ in the second term of equation \ref{uinv}) with
\begin{equation}
P^2 = \sqrt{\frac{(-\eta^2) E^2 B^2}{1-\frac{1}{\gamma_{max}^2}}}
\end{equation}
{\it only} when $1>\gamma^2>\gamma^2_{max}$.  This gives a limited
3-velocity of
\begin{equation}\label{vinvlimited}
v^i = -\beta^i + \alpha^2 \frac{[ijk] E_j B_k}{\detg P^2} ,
\end{equation}
that has a continuous transition between standard force-free
electrodynamics and dissipative electrodynamics.  The 4-velocity
given by equation (\ref{uinv}) is, by construction, limited to
always be time-like and have $\gamma\le\gamma_{max}$.  The
energy-momentum lost in such a limiting procedure is assumed to be
gained by inertial mass in the form of, e.g., (rapidly lost) thermal
or field-aligned kinetic energy that has no effect on the field.

Second, current sheets dissipate due to advection of field into the
sheet and subsequent reconnection and cancelation of the magnetic
field.  Even under exactly symmetric conditions, numerical round-off
error quickly leads to random velocity drifts across the current
sheet.  This magnetic field advection can be limited or avoided by
limiting the drift velocity perpendicular to the current sheet.  For
example, the 3-velocity given by equation (\ref{vinvlimited}) can be
further modified to have a small or vanishing component
perpendicular to the current sheet.  That is, if $n^i$ is the
spatial normal vector of the current sheet plane at a particular
time-slice, then we can set
\begin{equation}
n^j v^i g_{ij} = 0
\end{equation}
within some infinitesimal region bounding the current sheet.  This
changes Ohm's law in equation (\ref{Jexplain}) to have a vanishing
conductivity perpendicular to field lines, i.e. the spatial part of
$J_\perp^\alpha$ vanishes such that the
``$\mathbf{E}\times\mathbf{B}$ drift'' vanishes along the normal
direction of the current sheet. This explicitly avoids significant
reconnection by avoiding anomalous numerical drifts.  This is useful
in studying systems for which the effect of reconnection is
uncertain and the current sheet may be stable. The above
prescription can be generalized to set {\it any} arbitrary drift
speed into the current sheet.  Thus, physical models of the current
sheet and reconnection speed can be included in force-free models as
long as inertia plays no other role than in the sheet.

At present this is only implemented for a priori known locations of
the current sheet, although a general algorithm should be similar to
reconnection-capturing methods (see, e.g., \citealt{sp01}). In the
tests below that have a current sheet, there are 4 numerical grid
zones in a current sheet region that are forced to obey the above
condition.  This guarantees that the stencil, used by the
reconstruction and other dissipative procedures in HARM, does not
couple quantities diffusively across the sheet. This also ensures
that upon convergence testing that this numerical scheme to avoid
reconnection in current sheets plays no role in the results.

This formulation ensures that $B^2-E^2>0$ and that current sheets can
be forced to be mostly stable, unlike in \citet{kom02b,kras05} and is
improved compared to \citet{kom05b}.

\subsection{Quasi-GRMHD and Stationary Force-Free}\label{quasimhd}

While this paper describes a GRFFE formulation, since the same code
also has a GRMHD formulation, one can imagine hybrid schemes that
integrate the decoupled equations of motion in the stiff regime where
$b^2/\rho_0\gg 1$, where $\rho_0$ is the rest-mass density.  All GRMHD
numerical schemes have difficulties with this regime, so a decoupled
evolution may prove useful for studying the first nonzero order effect
of inertia in a force-free field.  This hybrid method will be
used in future work, but the method is outlined here.

In the force-free limit, one may still retain the evolution of the
rest-mass and internal energy density with no back reaction onto the
field (see also \citealt{ms94,cont95,cont99}).  The method is to
evolve the full GRMHD equations of motion, but to determine the
field-perpendicular velocity components and field from the force-free
equations alone.  In this case one can readily obtain the
field-aligned velocity component from the GRMHD inversion.  That is,
in ideal GRMHD in general, the fluid velocity may be broken into a
field-perpendicular and a field-aligned velocity or equivalently into
an electromagnetic and a matter velocity,
\begin{equation}
u^\alpha_{FL} = u^\alpha_\perp + u^\alpha_\parallel = u^\alpha_{EM} +
u^\alpha_{MA} .
\end{equation}

Also of interest is the fluid motion in a stationary force-free field,
which allows a study of the Lorentz factor along a force-free field
line \citep{ms94,cont95,cont99}.  For an stationary, axisymmetric
model the energy (Bernoulli) equation alone determines the
field-aligned velocity and the frozen-in conditions,
\begin{equation}
\frac{u^r_{FL}}{B^r} = \frac{u^\theta_{FL}}{B^\theta} = \frac{u^\phi_{FL} - \Omega_F
 u^t_{FL}}{B^\phi} ,
\end{equation}
apply to the fluid velocity and by $u^{EM}_\alpha B^\alpha=0$,
\begin{equation}\label{frozenffde}
\frac{u^r_{EM}}{B^r} = \frac{u^\theta_{EM}}{B^\theta} = \frac{u^\phi_{EM} - \Omega_F
 u^t_{EM}}{B^\phi} ,
\end{equation}
apply to the electromagnetic velocity.  These equations can be solved
for the electromagnetic 3-velocity to obtain
\begin{equation}\label{vem}
v^i_{EM} = \Omega_F \phi^i - B^i B_\mu \left(\frac{t^\mu + \Omega_F
\phi^\mu}{B^2}\right)
\end{equation}
where $t^\mu$ and $\phi^\mu$ are the time and $\phi$ Killing vectors
and $B_\mu = g_{\mu\nu} B^\nu$.  That is, for stationary,
axisymmetric solutions the magnetic field and $\Omega_F$ alone
determine the electromagnetic velocity, unlike in general as given
by equation (\ref{vinv}).

\section{Algorithm Tests}\label{tests}

\begin{figure*}
\includegraphics[width=6.6in,clip]{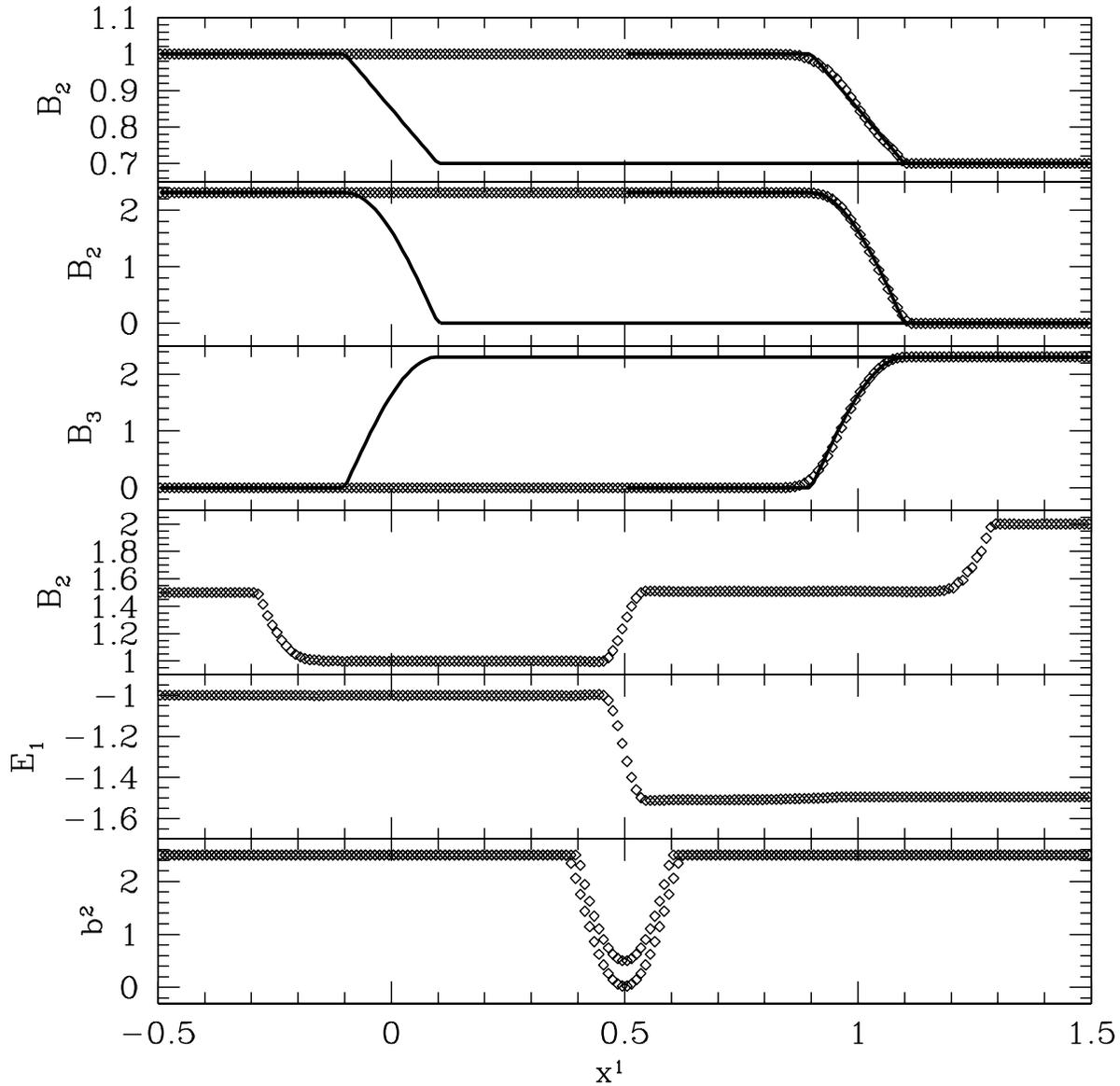}

\caption{Suite of tests described by \citet{kom02b}. Solid curves
  represent the initial and final analytic solution for the top three
  panels.  Diamonds represent the numerical solution at the final
  time, except for the plot of $b^2$ that shows both times.  Tests are
  as follows from top to bottom: 1) $B_2$ for fast wave ; 2) $B_2$ and
  $B_3$ for degenerate \alf~wave ; 3) $B_2$ and $E_1$ for three-wave
  problem ; and 4) initial and final smoothed $b^2=B^2-E^2\rightarrow
  0$ problem. }
\label{tests1}
\end{figure*}

A parameter space test of the analytic GRFFE inversion described above
was performed to evaluate the range of allowed Lorentz factors the
inversion is capable of handling.  The procedure is to start with a
known $v^i$ and $B^i$ for a specific point in space-time, determine
$T^t_\mu$, and then invert to get $v^i$.  Double precision is used for
all quantities.  The typical failure mode is that the velocity is
determined to be space-like, and this is nearly coincident with a
significant increase in the error in the inversion.  Of interest is
the general maximum value of the Lorentz factor below which no
failures occur.

In Minkowski space-time tests, the maximum Lorentz factor is
$u^t\sim 10^{10}$ before roundoff error causes the inversion to
suddenly have significant error.  Below this $u^t$, the relative
error is similar to machine error.  As an extreme test, the
space-time point is chosen in Kerr-Schild coordinates with a Kerr
spin parameter of $a/M=0.9375$ for a point on the horizon at
$\theta=\pi/4$.  For grid-aligned flows the maximum Lorentz factor
is $u^t\sim 10^{10}$ as before.  However, for arbitrary flow
directions, the inversion can fail for $u^t\gtrsim 2200$.  Most
astrophysical flows of interest have $u^t<2000$, so this is
sufficient for our purposes. Otherwise a smaller machine precision
should be used.

The GRFFE formulation is coupled to HARM to test the formulation's
ability to handle typical force-free problems.  The GRFFE formulation
is used in HARM to perform the Minkowski space-time test calculations
that are described by \citet{kom02b,kom04a}.  These tests include: 1)
a fast wave ; 2) a degenerate \alf~wave ; 3) a three-wave problem ; 4)
a problem that evolves to $B^2-E^2<0$ ; 5) a standing \alf~wave ; and
6) two current sheet problems.  Since our GRFFE formulation assumes
$\mathbf{E}\cdot\mathbf{B}=0$, their non-degenerate \alf~wave test is
not considered.

\begin{figure*}
\includegraphics[width=6.6in,clip]{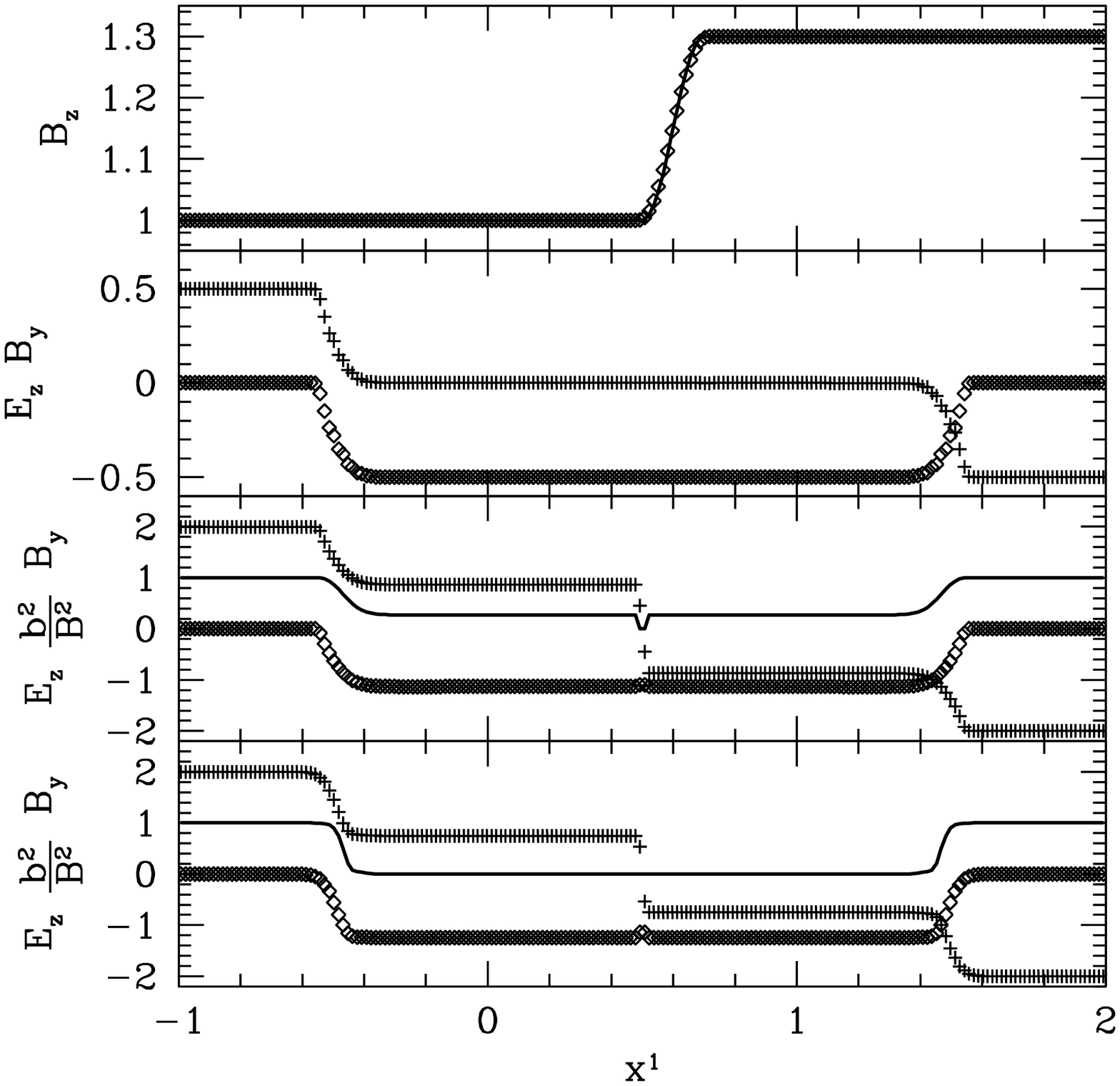}

\caption{Suite of tests described by \citet{kom04a}. Solid curves
  represent the initial and final analytic solution except in the
  bottom panel.  Pluses and diamonds represent the numerical solution
  at the final time.  Tests are as follows from top to bottom: 1)
  $B_z$ for stationary \alf~wave ; 2) $E_z$ and $B_y$ for current
  sheet with $B_0=0.5$ ; 3,4) $E_z$, $b^2$, and $B_y$ for current sheet
  with $B_0=2$.  Third is with 2nd order method and fourth is with 4th
  order method. }
\label{tests2}
\end{figure*}

Their notation of the ``wave frame'' fields $E'$ and $B'$ are
equivalent to the HARM notation for $e^\mu=0$ and $b^\mu$ for
$E'=0$, and otherwise the wave frame, $E'$, and $B'$ can be
processed through the inversion routine to setup an initial
lab-frame $B^i$ and $v^i$ from their $E'$, $B'$, and speed $\mu$.
The other problems they setup only require the lab-frame $E$ and
$B$, and again our GRFFE inversion can be used to obtain the initial
lab-frame $B^i$ and $v^i$ as long as the problem is degenerate. Note
that our sign conventions for $E$ and $B$ agree with theirs. We
estimate that they used about $200$ grid zones for their tests and
so use the same number for our tests.  We use the same final time,
box size, and plot labels for easy comparisons.  We shift the $x=0$
position to be similar for each set of tests in
\citet{kom02b,kom04a}.  A Courant factor of $0.9$ and the
monotonized central limiter is used for all tests except an
additional current sheet test with parabolic spatial interpolation.

Notice that apart from the initial condition given in
\citet{kom02b,kom04a}, for the fast wave one requires a relation
determined earlier in their paper: $E^z = C-\mu_f B^y$, where $C$ is
constrained such that $B^2-E^2<0$ ($C=1$ was chosen) and $\mu_f=+1$
was chosen by them.

Figure~\ref{tests1} shows the suite of tests demonstrated in
\citet{kom02b}.  For the top 3 panels, the solid line denotes the
analytic solution at the initial and final time.  The diamonds denote
the numerical solution at the final time.  The top panel shows $B_2$
for the fast wave.  The head of the wave is more resolved than in
\citet{kom02b}, but the tail of the wave is slightly less resolved.
The second and third panels from the top show $B_2$ and $B_3$ for the
degenerate \alf~wave problem.  The code does well to capture both fast
and \alf~waves.  The next two panels show $B_2$ and $E_1$ for the
three-wave problem.  The waves are resolved similarly as in
\citet{kom02b}.  The bottom panel shows $b^2/B^2=(B^2-E^2)/B^2$ for
the smoothed $B^2-E^2\rightarrow 0$ test problem.  Overall, the GRFFE
version of HARM performs comparably to the code by \citet{kom02b}
based on more complicated Riemann solvers.

Figure~\ref{tests2} shows the suite of tests demonstrated in
\citet{kom04a}.  The top panel shows $B_z$ for a stationary
\alf~wave. This wave is much more resolved than in \citet{kom04a}
and indicates that the code's effective resistive diffusion
coefficient is quite low.  The next panel shows $E_z$ and $B_y$ for
a current sheet problem as described in \citet{kom04a} with
$B_0=0.5$.  The features are well-resolved.  The two bottom panels
show the second current sheet problem with $B_0=2$.  The upper of
the two is using the MC limiter, while the lower of the two is using
a parabolic interpolation \citep{colella84}, which gives similar
results to \citet{kom04a}.  In the fast wave region the MC limiter
gives a Lorentz factor of $\Gamma=1.90$, while the parabolic method
gives $\Gamma=2000$, the largest allowed Lorentz factor.  For either
method, the region at $x^1=0.5$ within the current sheet reaches
$B^2-E^2\sim 0$, but the Lorentz factor limiter (here limited to
$\gamma_{max}=2000$) keeps the code stable despite the presence of
the current sheet.  No complicated dissipation model had to be
included to achieve such a result.

\section{Physical Models}\label{models}

This section considers astrophysical models for which the GRFFE
approximation is a reasonable one.  The GRFFE formulation is used in
HARM.

\citet{kom04a} study the Wald \citep{wald74} and split-monopole BZ77
\citep{bz77} solutions for slowly and moderately rapidly rotating
black holes. Of particular interest is whether the Wald and
split-monopole solutions can be better represented compared to as
shown in figures 3 and 4 of \citet{kom04a}.  We consider the
monopole and split-monopole for slowly rotating black holes in this
paper.

The other models are considered in a separate paper
\citep{mckinney05e}, which demonstrates our GRFFE formulation's
ability to handle the current sheet in the actual {\it
split}-monopole solution of \citet{bz77} and the Wald problem with a
rapidly rotating black hole spin.  In that paper for the Wald
problem, we were able to reach similar solutions \citet{kom05a} who
used an MHD model to avoid significant reconnection in the current
sheet that developed in their force-free models \citet{kom04a}.

Neutron star magnetospheres are studied in a separate paper
\citep{mckinney05d}, which demonstrates that the GRFFE code can be
used to study pulsar magnetospheres even in the presence of a current
sheet.  That paper shows we are able to avoid the problems encountered
by \citet{kom05b} with force-free and the current sheet.  We found
similar results to, but more accurate than, their ideal MHD results.

In this paper we focus on slowly rotating black hole magnetospheres
that require general relativity and so full GRFFE.  The interesting
quantities that are plotted throughout the following sections are
$\Omega_F \equiv F_{t\theta}/F_{\theta\phi}$, which is also
$\Omega_F = F_{tr}/F_{r\phi}$ for a stationary, axisymmetric flow.
This quantity often appears as a ratio to the black hole angular
velocity of $\Omega_H\equiv a/(2r_+)$.  The radial and $\theta$
magnetic field strengths for these plots is defined as in
\citet{kom04a}, with $B^i\equiv \dF^{it}$.  Also interesting is the
conserved toroidal flux of $\detg \tilde{B}_\phi\equiv B_\phi\equiv
\dF_{t\phi}=\detg F^{r\theta}$.\footnote{Notice the slight change,
for simplicity, in notation for $B_\phi$ and $B^i$ from this point
onward.} This is because the electromagnetic energy flux is $F^i_E =
-T^i_t = -B^i \Omega_F B_\phi$ and the electromagnetic angular
momentum flux is $F^i_L = F^i_E/\Omega_F$. Also of interest is the
magnetic vector potential ($A_\phi$), whose contours are plotted and
represent flow surfaces. For axisymmetric, stationary flows these
surfaces define surfaces of constant $\Omega_F$, $B_\phi$,
$F^i_E/B^i$, and $F^i_L/B^i$ in dissipationless force-free
electrodynamics.  Finally, also of interest are the light surfaces,
which are defined for the case of a purely rotational velocity
$\Omega=\Omega_F$ leads to a null trajectory, i.e. from $u^\alpha
u_\alpha=-1$,
\begin{equation}\label{null}
g_{tt} + 2\Omega g_{\phi t} + \Omega^2 g_{\phi\phi} = 0 .
\end{equation}

\subsection{Black Hole Magnetosphere: Blandford-Znajek Monopole and Split-Monopole }\label{bzsplitsection}

\citet{kom02b} demonstrated the stability of the {\it pure} monopole
solution of \citet{bz77} by studying one hemisphere.  Our GRFFE
formulation generates quantitatively similar results.  \citet{kras05}
also use a method based on HARM and study the dependence of the field
(purely monopolar-type) geometry and field rotation frequency on black
hole spin.  They also studied how the energy output varies with black
hole spin up to $a/M=0.9999$.  They find that the energy output
follows $\dot{E}\propto \Omega_H^2$ up to $a/M=0.98$.  This is in
contrast to the energy output in the presence of a thick disk that
follows $\dot{E}\propto \Omega_H^3\propto \Omega_F^3$ when accounting
for the mass accretion rate \citep{mckinney05a}.  This suggests that
the presence of matter can be important.  However, there may be
systems that are dominated by a magnetosphere rather than a disk (see,
e.g., \citealt{igumenshchev2003,narayan2003}).  For such systems, the
force-free limit may be sufficient to describe the energetics and
geometry of the field lines.

\begin{figure*}
\begin{center}
\includegraphics[width=4.0in,clip]{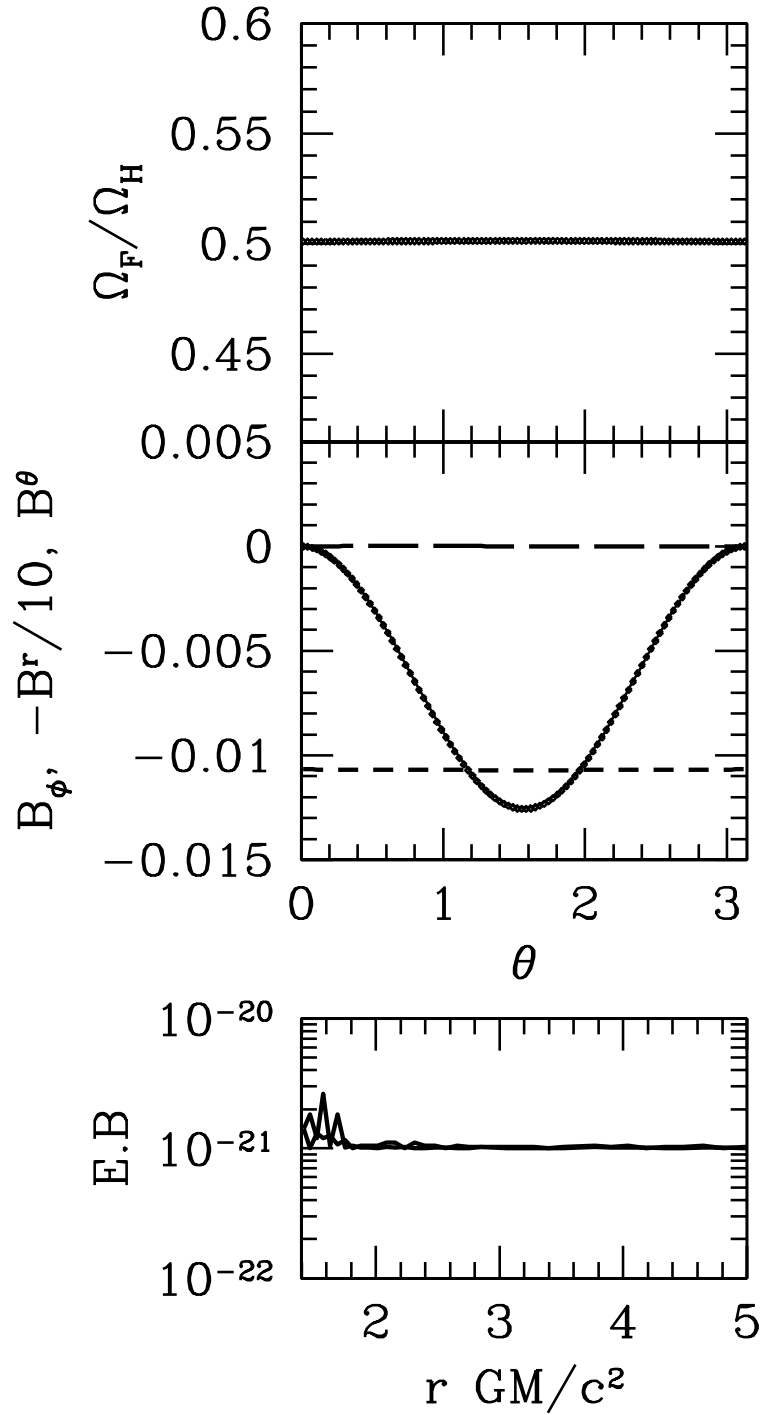}
\end{center}

\caption{Pure monopole Blandford-Znajek solution for black hole spin
$a/M=0.1$.  Upper panel: $\Omega_F/\Omega_H$.  Middle panel: $-B_\phi$
(dotted: numerical, solid: analytic), $B^r$ (short dashed), and
$B^\theta$ (long dashed).  Lower panel:
$\mathbf{E}\cdot\mathbf{B}$. Model at $t=50GM/c^3$.  Directly
comparable with figure 2 in \citet{kom04a}.  }
\label{bzpure}
\end{figure*}

For the purposes of testing, the pure monopole Blandford-Znajek
solution is considered, as in section 5.2 of \citet{kom04a}.  An
identical numerical setup to \citet{kom04a} is used.  A grid is chosen
with an outer radius of $r=260GM/c^2$.  There are $150$
equally-logarithmically-spaced radial zones and $100$ uniform $\theta$
zones.  As mentioned in \citet{kom04a}, the final result is
insensitive to the details of the initial conditions, so the simple
$a/M=0$ monopole solution is chosen as an initial condition.  The
evolution proceeds until $t=50GM/c^3$.

Figure~\ref{bzpure} shows $\Omega_F/\Omega_H$ (top panel), $-B_\phi$
(middle panel; pluses), analytic Blandford-Znajek (BZ) solution for
$-B_\phi$ (middle panel; solid line), $-B^r/10$ (middle panel; short
dashed line), $B^\theta$ (middle panel; long dashed line), and
$\mathbf{E}\cdot\mathbf{B}$ for $\theta=\pi-0.1$ and $\theta=0.1$.
The figure is directly comparable to figure 2 in \citet{kom04a}.  The
value of $\Omega_F/\Omega_H$ only varies monotonically from $0.5013$
near the equator to $0.5009$ at the poles.  Unlike in \citet{kom04a},
the value does not rise again near the poles, which suggests HARM is
properly resolving the coordinate singularity.  Their value is up to
$0.503$ near the equator, which suggest HARM is doing marginally
better.  The middle panels shows that the code is very accurately
reproducing the BZ solution.  The solid line indicating the analytic
solution is not visibly deviating from the crosses that mark the
numerical solution.  The bottom panel just advertises the fact that
HARM is uniformly preserving $\mathbf{E}\cdot\mathbf{B}=0$ to machine
precision, where figure 2 in \citet{kom04a} shows nonuniform errors of
order $10^{-4}$.

\begin{figure*}
\begin{center}
\includegraphics[width=5.0in,clip]{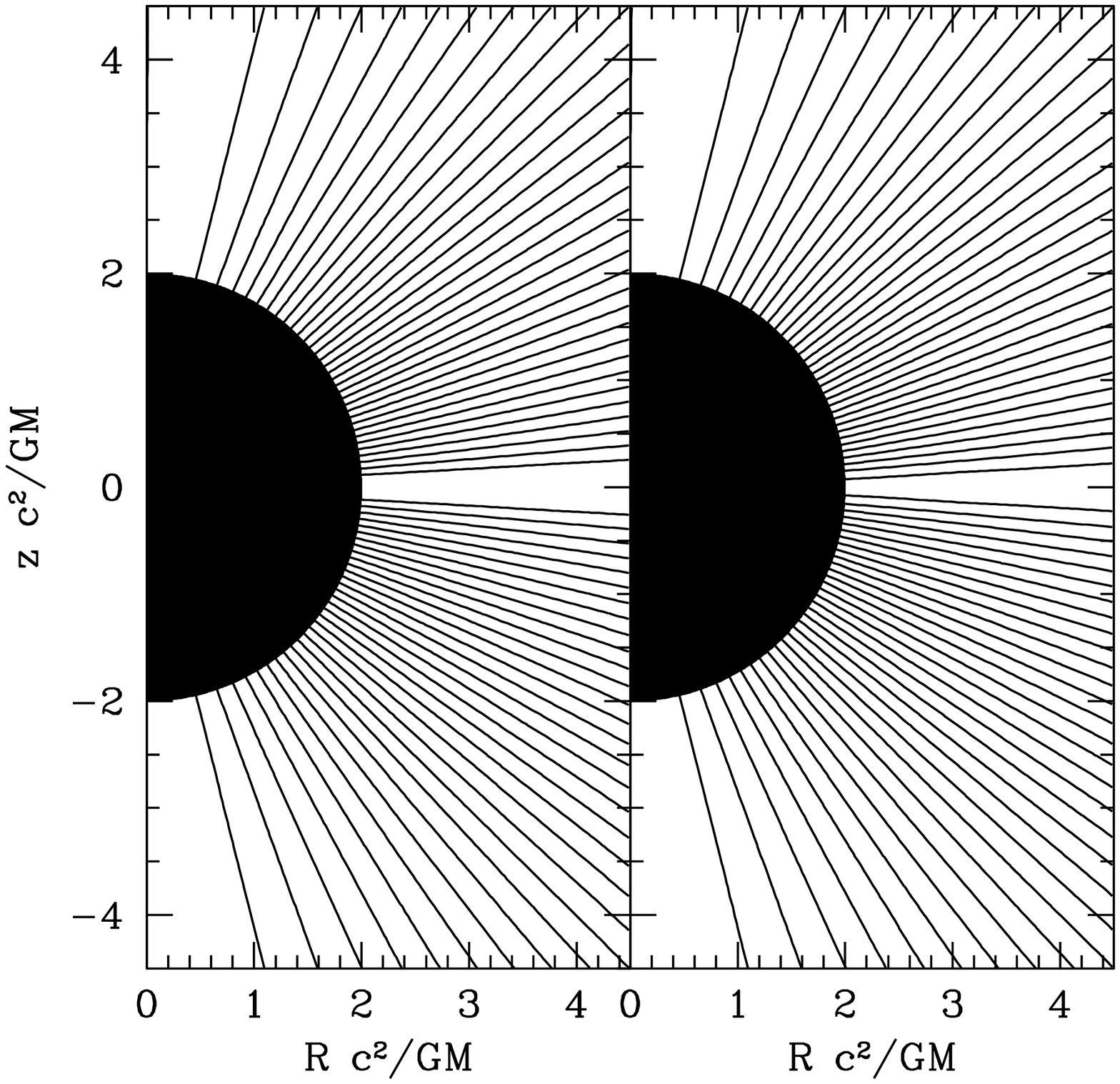}
\end{center}
\caption{ Contours of the magnetic vector potential component
$A_\phi$, giving the field lines for the split-monopole
Blandford-Znajek solution for black hole spin $a/M=0.1$.  Left panel:
Initial Model.  Right panel: Model at $t=5GM/c^3$.  Directly
comparable with figure 3 in \citet{kom04a}.  }
\label{bzsplit}
\end{figure*}

\citet{kom04a} study the Wald and split-monopole BZ77 solution for
slowly and moderately rapidly rotating black holes. Of particular
interest is whether the Wald and split-monopole solutions can be
better represented compared to as shown in figures 3 and 4 of
\citet{kom04a}.  This paper demonstrates our GRFFE formulation's
ability to handle the current sheet in the actual {\it
split}-monopole solution of \citet{bz77}.

We use an identical setup to \citet{kom04a} for the split-monopole
case with a black hole with spin $a=0.1$\footnote{\citet{kom04a}
pointed out in their section 5.3 that $a=0.1$, but then their figure 3
caption says ``Schwarzschild''.  The $a=0$ and $a=0.1$ models show the
same results, so this issue is not crucial.}.  There are $100$ uniform
$\theta$ zones with $80$ radial zones such that $dr/r={\rm Const.}$.
The model is axisymmetric and the inner boundary is placed inside the
horizon and the outer boundary is at $r=29GM/c^2$.  The model is run
till $t=5GM/c^3$ as in \citet{kom04a}.

Figure~\ref{bzsplit} shows a two panel plot that can be directly
compared with figure 3 in \citet{kom04a}.  In their model the
current sheet rapidly reconnects even over this very short time
period.  Our results show essentially zero reconnection.  This is
primarily due to the modifications of the drift velocity described
in section~\ref{diss}, and secondarily due to the low diffusivity of
the algorithm as demonstrated in section~\ref{tests}.  Without the
drift velocity modification, however, our results are similar to
shown in figure 3 of \citet{kom04a}.  So this demonstrates the
usefulness of the relatively ad hoc approach of treating current
sheets.

For this model, we used the local Lax-Friedrich (LLAXF), rather than
the Harten-Lax-van Leer (HLL), approximate nonlinear Riemann solver
(see, e.g. \citealt{gmt03}).  The HLL solver generated fluctuations at
the inner-radial boundary that back propagate through the horizon due
to numerical diffusion.  HLL is explicitly causal. However, in HARM
and all numerical methods the author is aware of (for an exception
see, e.g. \citealt{hawke05}), the stencil used to reconstruct the
quantities to the Riemann interface is still acausal.  For example, in
HARM, the stencil is always centered and so connects information
across the horizon depending upon the smoothness of the flow.  The
author knows of no numerical method that accounts for the
characteristic information in constraining the reconstruction to be
causal.  In \citet{mckinney05e}, we consider modifications to the
stencil that enforce strict causality.  This eliminates all the
problems described above.

\section{Conclusions}\label{conclusions}

We described a GRFFE formulation that allows GRMHD codes to directly
evolve the GRFFE equations of motion.  Rather than evolving the
electric and magnetic fields, the velocity and magnetic field are
directly evolved.  This formulation strictly enforces
$\mathbf{E}\cdot\mathbf{B}=0$, $\nabla\cdot\mathbf{B}=0$, and energy
conservation, unlike prior codes.  Established, accurate, and
well-tested GRMHD codes can simply add a new inversion piece of code
to their existing code, while continuing to use all the
already-developed facilities present.

We also introduced a simplified general model of the dissipation of
the electric field to enforce the $B^2-E^2>0$ constraint.  This limits
the code to a regime of Lorentz factors that the code can handle
without significant numerical errors.

A simplified general model was introduced to allow current sheets to
be resolved, without reconnection, over many dynamical times.  The
other improvements to the code, such as strict enforcement of
$\mathbf{E}\cdot\mathbf{B}=0$, $\nabla\cdot\mathbf{B}=0$, energy
conservation, and $B^2-E^2>0$ do not play a role in this improvement.
Limiting the numerically induced drift velocity perpendicular to the
sheet was a crucial step to make such a force-free code useful, and
resolves the difficulties encountered by prior authors
\citep{kom04a,kom05b}.  For highly magnetized systems, our GRFFE
results are as good as ideal MHD results without the need to introduce
an artificial evolution of the rest-mass and internal energy
densities.  Evidence of this fact is shown in other papers
\citep{mckinney05d,mckinney05e}, where we model particular
astrophysical systems.

Numerical tests showed that the GRFFE formulation as used in HARM is
robust and accurate.  HARM uses a simplified nonlinear approximate
Riemann solver as in \citet{kras05}, which, while being much simpler
than the exact Riemann solution by \citet{kom02b}, produced as
accurate results.

For the pure monopole Blandford-Znajek model, we found similar
results as \citet{kom02b,kras05}.  We also demonstrated our code's
ability to handle current sheets, which was found to be difficult in
\citet{kom04a}.  Here and in \citet{mckinney05e}, the split-monopole
problem was able to be solved without significant reconnection. This
allows one to study general magnetospheres with arbitrary currents
in a sheet, such as the paraboloidal field. \citet{mckinney05e} also
discusses the Wald solution, and it shows that our code is able to
obtain similar results as the MHD code of \citet{kom05a}, while
\citet{kom04a} encountered difficulties with the current sheet
reconnecting too fast.

\section*{Acknowledgments}

This research was supported by NASA-Astrophysics Theory Program grant
NAG5-10780 and a Harvard CfA Institute for Theory and Computation
fellowship.  I thank Alexandre Tchekhovskoi and Avery Broderick for
useful discussions that pushed the project forward.  I thank Ian
Morrison, Ruben Krasnopolsky, and Charles Gammie for discussions
regarding the original simplified HARM-like GRFFE formulation during
my time with them at the UIUC.  I thank Serguei Komissarov and Dmitri
Uzdensky for comments.


\appendix




\label{lastpage}

\end{document}

%% file: macros.tex
\newcommand{\detg}{{\sqrt{-g}}}

\newcommand{\dF}{{^{^*}\!\!F}}
\newcommand{\alf}{Alfv\'en}

\def\fps@figure{bp}%
\def\fps@table{bp}%
\def\fps@plate{bp}%
\def\eps@scaling{1.0}%
\newcommand\epsscale[1]{\gdef\eps@scaling{#1}}%
\newcommand\plotone[1]{%
 \centering
 \leavevmode
 \includegraphics[width={\eps@scaling\columnwidth}]{#1}%
}%
\newcommand\plottwo[2]{%
 \centering
 \leavevmode
 \columnwidth=.45\columnwidth
 \includegraphics[width={\eps@scaling\columnwidth}]{#1}%
 \hfil
 \includegraphics[width={\eps@scaling\columnwidth}]{#2}%
}%
\newcommand\plotfiddle[7]{%
 \centering
 \leavevmode
 \vbox\@to#2{\rule{\z@}{#2}}%
 \includegraphics[%
  scale=#4,
  angle=#3,
  origin=c
 ]{#1}%
}%

\newcommand\apj{\rmfamily{ApJ}}%
\newcommand\apjl{\rmfamily{ApJ}}%
\newcommand\aap{\rmfamily{A\&A}}%
\newcommand\mnras{\rmfamily{MNRAS}}%
\newcommand\prd{\rmfamily{Phys.~Rev.~D}}%
\newcommand\pasj{\rmfamily{PASJ}}%
\let\la=\lesssim            
\let\ga=\gtrsim